\documentclass[aip,jcp,reprint,showkeys]{revtex4-1}
\usepackage{bm,graphicx,tabularx,array,booktabs,dcolumn,xcolor,microtype,multirow,amscd,amsmath,amssymb,amsfonts,physics,siunitx,tikz,wrapfig,enumitem,float}
\usepackage[version=4]{mhchem}
\usepackage[utf8]{inputenc}
\usepackage[T1]{fontenc}
\usepackage{txfonts}
\usepackage[normalem]{ulem}
\newcommand{\mc}{\multicolumn}

\usepackage{tikz}
\usetikzlibrary{positioning}

\usepackage[
	colorlinks=true,
    citecolor=blue,
    linkcolor=blue,
    filecolor=blue,      
    urlcolor=blue,	
    breaklinks=true
	]{hyperref}
\newcommand{\red}[1]{\textcolor{red}{#1}}
\renewcommand{\red}[1]{#1}

\usepackage{caption} 
\usepackage{subcaption}

\newcommand{\kB}{k_\text{B}}
\newcommand{\rc}{r_\text{c}}

\newcommand{\lbp}{\lambda_{\text{bp}}}

\newcommand{\e}{\mathrm{e}}
\renewcommand{\i}{\mathrm{i}}
\renewcommand{\H}{H}
\newcommand{\Href}{\H^{(0)}}
\newcommand{\Hper}{\H^{(1)}}

\newcommand{\UOX}{Physical and Theoretical Chemistry Laboratory, Department of Chemistry, University of Oxford, South Parks Road, Oxford, OX1 3QZ, U.K.}

\begin{document}

\title{Complex analysis of divergent perturbation theory at finite temperature}

\author{Yi~Sun}
\affiliation{\UOX}

\author{Hugh~G.~A.~Burton}
\email{hugh.burton@chem.ox.ac.uk}
\affiliation{\UOX}

\date{\today}

\begin{abstract}
We investigate the convergence properties of finite-temperature perturbation theory by considering the mathematical structure of thermodynamic potentials using complex analysis. We discover that zeros of the partition function lead to poles in the internal energy and \red{logarithmic singularities in the Helmholtz free energy} which create divergent expansions in the canonical ensemble. Analysing these zeros reveals that the radius of convergence increases for higher temperatures. 
\red{In contrast, when the reference state is degenerate, these poles in the internal energy create a zero radius of convergence in the zero-temperature limit.}
Finally, by showing that the poles in the internal energy reduce to exceptional points in the zero-temperature limit, we unify the two main mathematical representations of quantum phase transitions.
\end{abstract}

\maketitle
\raggedbottom

Thermodynamic effects in electronic structure theory become significant when the 
band gap is comparable to the temperature. 
This scenario arises under extreme conditions such as planetary interiors
and laser pulses,\cite{GrazianiBook}
or for systems with low-energy  excitations including metals and semiconductors.\cite{Kotliar2006}
Finite-temperature effects also play a role in the emergence of 
quantum phase transitions, with applications in many-body localisation, 
magnetic phases, and high-temperature superconductivity.\cite{Kotliar2006,SachdevBook}

Perturbation theory is the most established \textit{ab initio} approach for finite-temperature 
systems where electron correlation effects are important.\cite{MattuckBook,MarchBook,Jha2020,Hirata2021b}
However, for low-order perturbation theory to be reliable and systematically improvable, 
the corresponding expansion should give a convergent series.
Zero-temperature perturbation theory can become divergent
when the reference state is a poor approximation to the physical system,  or when 
there are near-degeneracies in the reference energies [see Ref.~\onlinecite{Marie2021} for a review].
In contrast, energy degeneracies are less significant at 
finite temperature  than in the zero-temperature expansion,\cite{Santra2017}
\red{although the series convergence can often worsen for lower temperatures.
Furthermore,  finite-temperature perturbation theory can diverge at zero temperature even 
when the zero-temperature perturbation expansion is convergent.\cite{Hirata2021}
This situation arises for degenerate or incorrect reference states and its manifestation in electronic perturbation 
theory is known as the Kohn--Luttinger problem.\cite{Kohn1960}}
However, the general relationship between the convergence of finite-temperature perturbation 
theory and its zero-temperature counterpart has not yet been fully established.

The convergence of a perturbation expansion $\H=\Href+\lambda \Hper$ 
can be mathematically examined by investigating the structure of the 
energy function $E(\lambda)$ in the complex-$\lambda$ plane.
From complex analysis, the radius of convergence $\rc$ for a perturbation expansion of a function $f(\lambda)$
is determined by the distance of the closest singularity of $f(\lambda)$ 
to the origin in the complex-$\lambda$ plane.\cite{BenderBook,Goodson2012,Stillinger2000,Sergeev2005}
These singularities represent points where $f(\lambda)$ becomes non-analytic 
and may correspond to a pole, a branch point,
or a more complicated non-analytic feature.\cite{Sergeev2005,Stillinger2000}
The perturbation expansion will converge for the physical system at $\lambda = 1$ when $\rc > 1$ and will diverge for $\rc < 1$.
Therefore, understanding the convergence of finite-temperature perturbation theory requires
a detailed investigation into the structure of thermodynamic functions in the complex-$\lambda$ plane.

Complex analysis also plays an important role in the theory of thermodynamic phase transitions.
In Lee--Yang theory, zeros of the partition function exist at complex temperatures for finite systems
near a phase transition.\cite{Yang1952,Lee1952}
These zeros converge onto the real axis in the thermodynamic limit and intersect at the 
critical temperature.
The same phenomenon occurs for any complex-valued control parameter,\cite{FisherBook} allowing Lee--Yang theory 
to be applied to zero-temperature quantum phase transitions.\cite{Kist2021}
Alternatively, avoided level crossings in finite systems are related to non-Hermitian exceptional points, where two energy levels 
become identical for a complex control parameter.\cite{Heiss2005,Heiss2012}
The distance of an exceptional point to the real-axis controls the ``sharpness'' of the avoided crossing and, in the 
thermodynamic limit of a quantum phase transition, the exceptional points converge onto the real axis.\cite{Heiss1990}
These exceptional points also play a pivotal role in the convergence of zero-temperature 
perturbation theory,\cite{Marie2021,Heiss2012}
symmetry breaking in mean-field approximations,\cite{Burton2019a,Burton2021b} and the convergence of 
quantum criticality in the complete basis set limit.\cite{Kais2006}

\begin{figure*}[htb!]
\begin{subfigure}{0.24\textwidth}
\includegraphics[width=\linewidth]{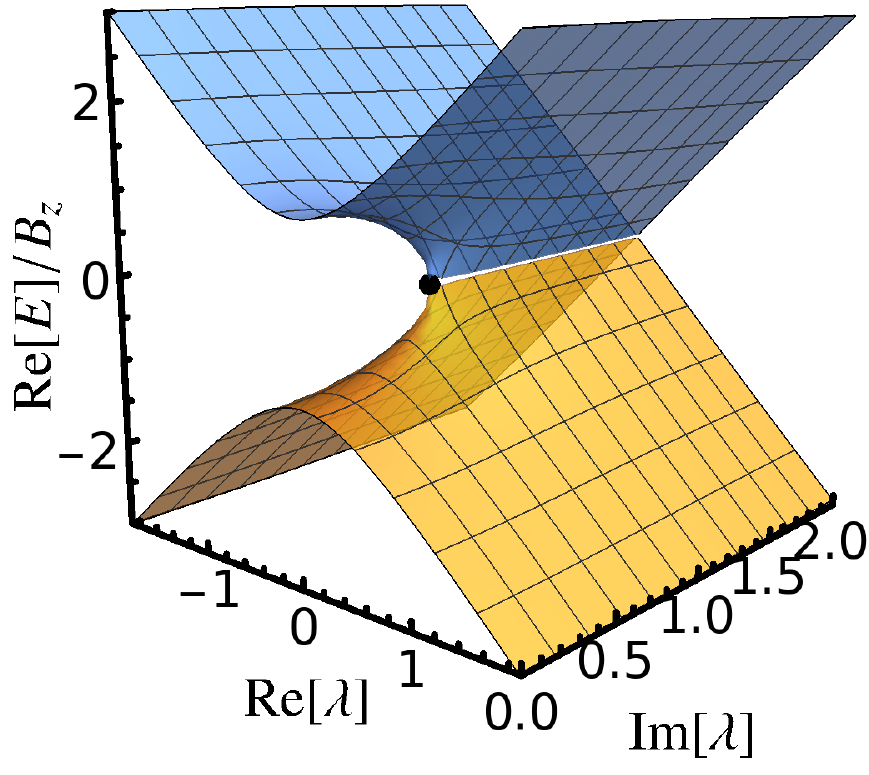}
\subcaption{$\kB T=0.00$ \label{subfig:T0.00}}
\end{subfigure}
\begin{subfigure}{0.24\textwidth}
\includegraphics[width=\linewidth]{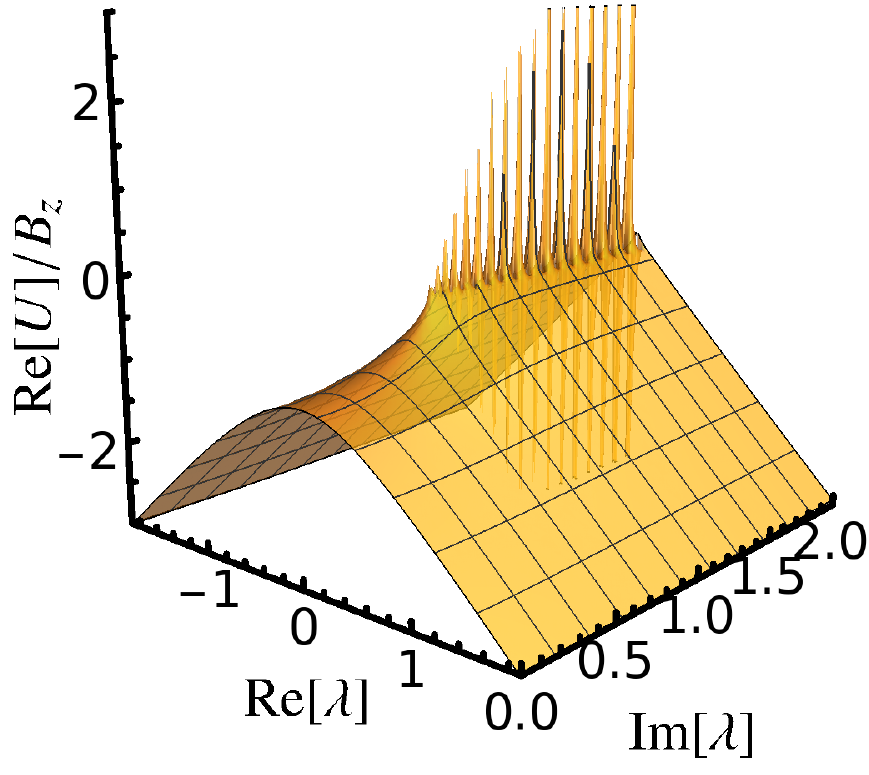}
\subcaption{$\kB T=0.05$ \label{subfig:T0.05}}
\end{subfigure}
\begin{subfigure}{0.24\textwidth}
\includegraphics[width=\linewidth]{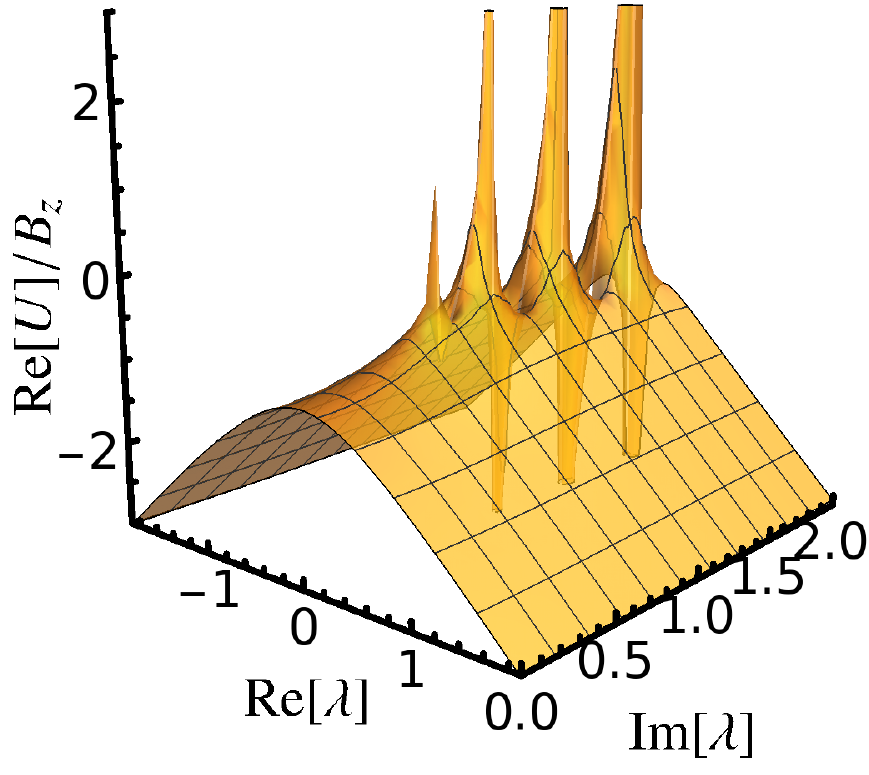}
\subcaption{$\kB T=0.25$ \label{subfig:T0.25}}
\end{subfigure}
\begin{subfigure}{0.24\textwidth}
\includegraphics[width=\linewidth]{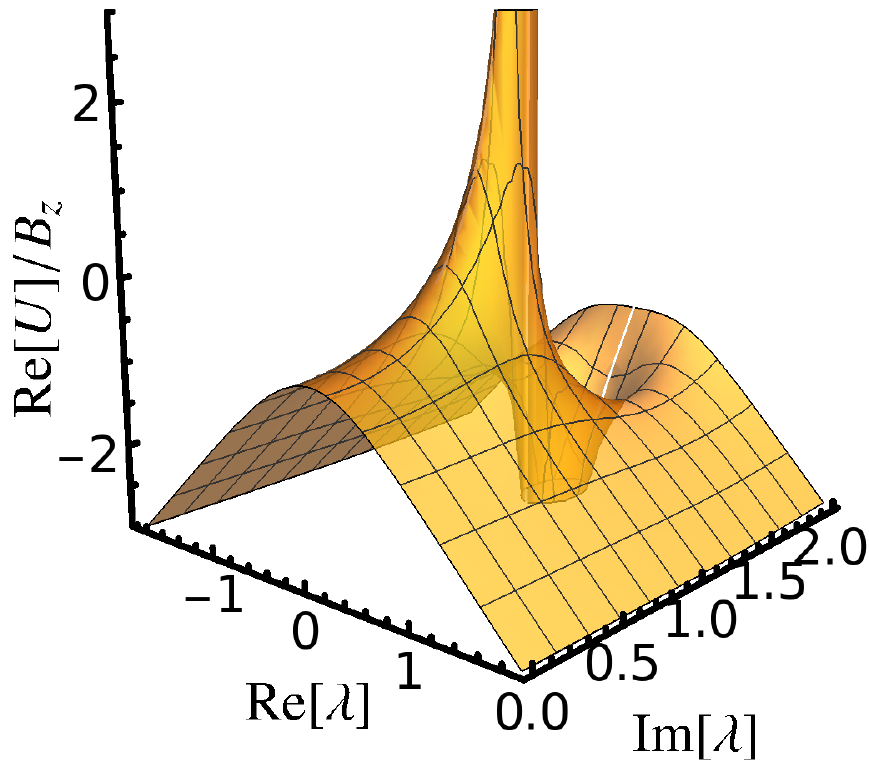}
\subcaption{$\kB T=1.00$ \label{subfig:T1.00}}
\end{subfigure}
\caption{Internal energy for the two-level Hamiltonian [Eq.~\eqref{eq:TLSHam}] in the canonical
ensemble as a function of the perturbation strength $\lambda$ for $B_x = \frac{3}{2} B_z$.
(a) The zero-temperature ground and excited state energies form a two-sheeted Riemann
surface with square-root branch points at $\lambda = \pm \mathrm{i}\, B_z / B_x$ (black dot).
(b)--(d) At finite temperature, the internal energy features a sequence of poles corresponding to 
zeros of the partition function $Z(\lambda)$, with the spacing increasing at higher temperature. 
The distance of the closest pole to the origin 
determines the radius of convergence for a perturbation expansion of the internal energy.
}
\label{fig:TLSenergy}
\end{figure*}

In this work, we investigate the convergence of finite-temperature perturbation theory
in the canonical ensemble through the lens of complex analysis. 
We find that the internal energy and Helmholtz free energy are punctuated by poles or 
logarithmic singularities in the complex-$\lambda$ plane
that determine the radius of convergence of the perturbation series.
These singularities are created by Lee--Yang zeros of the partition function and move further 
from the origin as the temperature increases.
Consequently, perturbation expansions converge increasingly rapidly at higher temperatures, even for a divergent 
zero-temperature expansion.
\red{For a degenerate reference, the zeros of the partition function converge onto $\lambda =0$ for $T\rightarrow0$, 
creating a mathematically undefined perturbation expansion that is the origin of the Kohn--Luttinger problem.} 
Finally, we extend these results to directly connect zero-temperature exceptional points and finite-temperature 
zeros of the partition function in the theory of quantum phase transitions.

We illustrate these ideas using a two-level system representing a spin-1/2 particle in 
a magnetic field. 
In the spinor basis of the $\hat{S}_z$ operator, the Hamiltonian is 
\begin{equation}
\H =
\begin{pmatrix}
-B_z & -B_x \\ 
-B_x & B_z
\end{pmatrix},
\label{eq:SpinHam}
\end{equation}
where $B_z$ and $B_x$ are the (real) components of the magnetic field along the $z$ and $x$
directions respectively.
For $B_x=0$ and $B_z > 0$, the ground state represents an electron aligned in the positive $z$ direction while 
the excited state represents an electron aligned with the negative $z$ direction.
In the $B_z \rightarrow 0$ limit, the electron in its ground state 
aligns with the positive (negative) 
$x$ direction for positive (negative) $B_x$ and a quantum phase transition occurs at $B_x = 0$.
Any two-state system of this type can be seen as a qubit. 



Within perturbation theory, the Hamiltonian is partitioned into a Hermitian reference Hamiltonian $\Href$ and a perturbation
$\Hper$.\cite{BenderBook}
The time-independent Schr\"{o}dinger equation is then recast as
\begin{equation}
\H(\lambda) \Psi_k(\lambda) = \qty(\Href + \lambda \Hper) \Psi_k(\lambda) = E_k(\lambda) \Psi_k(\lambda),
\end{equation}
where the parameter $\lambda$ controls the strength of the perturbation.
The exact energies and wave functions of the ground and excited states become $\lambda$-dependent functions, 
with $\lambda = 0$ corresponding to the reference model and $\lambda = 1$ representing the physical system.
Expanding the energy around $\lambda = 0$  gives the power series 
\begin{equation}
E_k(\lambda) = \sum_{j=0}^{\infty} E_k^{(j)} \lambda^{j}, 
\label{eq:EnergyPT}
\end{equation}
where $E_k^{(j)}$ provides the $j$th-order perturbation correction.
This expansion has a radius of convergence $\rc$ that controls
the values of $\lambda$ where the partial sums of increasing length tend towards the exact value of $E_k(\lambda)$.

When $\H(\lambda)$ is considered in the complex-$\lambda$ plane, it becomes non-Hermitian 
and the discrete eigenvalues become unified as a continuous Riemann surface.\cite{BenderPTBook} 
This Riemann surface represents a `one-to-many' function with each sheet representing a different eigenstate.
The most common singularities on $E(\lambda)$ are exceptional points (or branch points) where two energy levels 
become degenerate and the eigenstates become identical.\cite{BenderBook,Heiss2012,MoiseyevBook,Olsen1996,Olsen2000,Goodson2012}
These non-Hermitian features are related to the onset of dynamic stabilities, avoided level crossings,
and quantum phase transitions.\cite{Heiss2012}
Remarkably, following an eigenstate around an exceptional point interconverts the two 
energy levels;\cite{BenderPTBook,MoiseyevBook,Berry2011,Burton2019a} 
a property that has even been realised experimentally.\cite{Bittner2012,Doppler2016}

The reference and perturbation Hamiltonians for the two-level system can be defined as
\begin{equation}
\Href=
\begin{pmatrix}
-B_z & 0 \\ 
0 & B_z
\end{pmatrix}
\quad\text{and}\quad
\Hper=
\begin{pmatrix}
0 & -B_x \\ 
-B_x & 0 
\end{pmatrix},
\label{eq:TLSHam}
\end{equation}
with the exact zero-temperature eigenstates 
\begin{align}
E_{\pm}(\lambda) = \pm \sqrt{B_z^2 + \lambda^2 B_x^2}.
\label{eq:TLenergy}
\end{align}
\red{Here, the ground and excited state are defined as $E_{1} = E_{-}$ and $E_{2} = E_{+}$ respectively}.
These eigenstates form an artificial avoided level crossing along the real-$\lambda$ axis at $\lambda = 0$ 
that mirrors the true avoided crossing along the real $B_x$ axis.
The corresponding two-sheeted Riemann surface is shown in Figure~\ref{subfig:T0.00}.
When the discriminant in Eq.~\eqref{eq:TLenergy} is zero, the eiegenstates become degenerate,
creating square-root branch points at $\lbp = \pm \mathrm{i}\, B_z / B_x$.
Since these exceptional points are the only singularities in $E(\lambda)$, the radius of convergence is $\rc = \abs{\frac{B_z}{B_x}}$
and the zero-temperature perturbation expansion converges at the physical value $\lambda =1$
when $\abs{B_x} < \abs{B_z}$.

%
We now consider a canonical ensemble of non-interacting subsystems, where the partition function of 
a general $N$-level system is
\begin{equation}
Z(\lambda) = \sum_{k=1}^{N} \exp(-\beta E_k(\lambda)).
\end{equation}
Here, $\beta = (\kB T)^{-1}$ and $\kB$ and $T$ denote the Boltzmann constant and temperature respectively.
The internal energy is
\begin{equation}
U(\lambda) = - \frac{\dd \ln Z(\lambda)}{\dd \beta} = \frac{\sum_{k=1}^{N} E_k(\lambda) \exp(-\beta E_k(\lambda))}{Z(\lambda)},  
\label{eq:Udef}
\end{equation}
and the Helmholtz free energy is
\begin{equation}
F(\lambda) = -\frac{1}{\beta}\ln Z(\lambda).
\label{eq:Fdef}
\end{equation}
The perturbation series for these thermodynamic potentials are given by 
Taylor expansions around $\lambda = 0$,\cite{Jha2020} i.e.
\begin{equation}
U(\lambda) = \sum_{j=0}^{\infty} U^{(j)} \lambda^{j}
\quad\text{and}\quad
F(\lambda) = \sum_{j=0}^{\infty} F^{(j)} \lambda^{},
\end{equation}
where the individual corrections are, e.g.
\begin{equation}
U^{(j)} =\frac{1}{j!} \left. \frac{\partial^j U(\lambda) }{ \partial \lambda^j} \right|_{\lambda=0}.
\label{eq:def_derivates}
\end{equation}
The radius of convergence is then controlled by the distance of the closest singularity to the origin
in the complex-$\lambda$ plane.

Singularities in $U(\lambda)$ at non-zero temperature can be 
identified by inspecting the general form of Eq.~\eqref{eq:Udef}.
One possibility is that $U(\lambda)$ \red{has a singularity} at values of $\lambda$ corresponding 
to exceptional points in the zero-temperature energy.
We find that this is not the case because a thermodynamic summation including every discrete eigenstate
removes \red{singularities caused by exceptional points}. 
This property is shown for the partition function in Appendix~\ref{apdx:AnalyticU} 
and can be extended to the numerator of the internal energy.

On the other hand, $Z(\lambda)$ can become zero in the complex-$\lambda$ plane, creating poles 
in the internal energy that dictate the radius of convergence for the perturbation expansion.
At large temperatures, we find $Z(\lambda) \rightarrow 1$ for all $\lambda$. 
However, if we consider low temperatures near a degeneracy between $E_1(\lambda)$ and $E_2(\lambda)$,
the partition function approximates to
\begin{equation}
Z(\lambda) \approx \e^{-\beta E_1 (\lambda)} \qty[1 + \e^{ -\beta( E_2(\lambda) - E_1(\lambda))} ].
\end{equation}
Zeros of $Z(\lambda)$ then occur when $\e^{ -\beta( E_2(\lambda) - E_1(\lambda))} = -1$, giving an infinite 
set of possible solutions that satisfy
\begin{equation}
E_2(\lambda) - E_1(\lambda) = \underbrace{\kB T (2n+1) \pi}_{\omega_n}\, \i \quad \forall \quad n \in \mathbb{Z}.
\label{eq:Locus}
\end{equation}
In other words, zeros of the partition function for low-$T$ occur when the real components of $E_1(\lambda)$ and $E_2(\lambda)$ 
are degenerate and the difference in the imaginary components is equal to the Matsubara frequencies $\omega_n$.\cite{Matsubara1955}
For $T\rightarrow 0$, the Matsubara frequencies form a continuum along the imaginary axis and
zeros of the partition function occur whenever the real components are degenerate, regardless of the imaginary components.
\red{In this limit, the zeros of the partition function form a continuum along the locus of points where 
the real component of the lowest excitation energy is zero.
These points include non-Hermitian branch cuts on the ground-state energy surface as well as Hermitian conical intersection seams.}
The same zeros of $Z(\lambda)$ create logarithmic singularities in the Helmholtz free energy, meaning that 
$F(\lambda)$ and $U(\lambda)$ have an identical radius of convergence.

In the two-level system, the internal energy is given in terms of the perturbation strength $\lambda$ as
\begin{align}
U(\lambda) &= - \frac{\sqrt{B_z^2 + \lambda^2 B_x^2}\, \sinh(\beta \sqrt{B_z^2 + \lambda^2 B_x^2})}{\cosh(\beta \sqrt{B_z^2 + \lambda^2 B_x^2})},
\label{eq:Utls}
\end{align}
where the denominator represents the partition function $Z(\lambda)~=~\cosh(\beta \sqrt{B_z^2 + \lambda^2 B_x^2})$.
Plotting the internal energy at various temperatures in Figures~\ref{subfig:T0.05}--\subref{subfig:T1.00}, we find that the square-root branch cut in the zero-temperature
energy is lost for $T\neq 0$.
Instead, there are a sequence of poles with a separation that increases for higher temperatures.
Solving $Z(\lambda) = 0$, the positions of these poles is
\begin{equation}
\lambda_\text{pole} = \pm \mathrm{i} \frac{B_z}{B_x} \sqrt{1 + \frac{\pi^2 \kB^2 T^2 }{4{B_z}^2} (4n+1)^2} \quad \forall \quad n \in \mathbb{Z}.
\end{equation}
Therefore, the poles extend along the imaginary axis and their spacing 
decreases with temperature.
When $T\rightarrow0$, the poles tend towards a continuum extending outwards from the zero-temperature exceptional points, \red{corresponding to the line where
the real components of the ground- and excited-state energies are degenerate.
Ultimately, this continuum of poles recovers the branch cut on the lowest-energy sheet of the zero-temperature Riemann surface [Figure~\ref{subfig:T0.00}].}

\begin{figure}[b!]
\includegraphics[width=\linewidth]{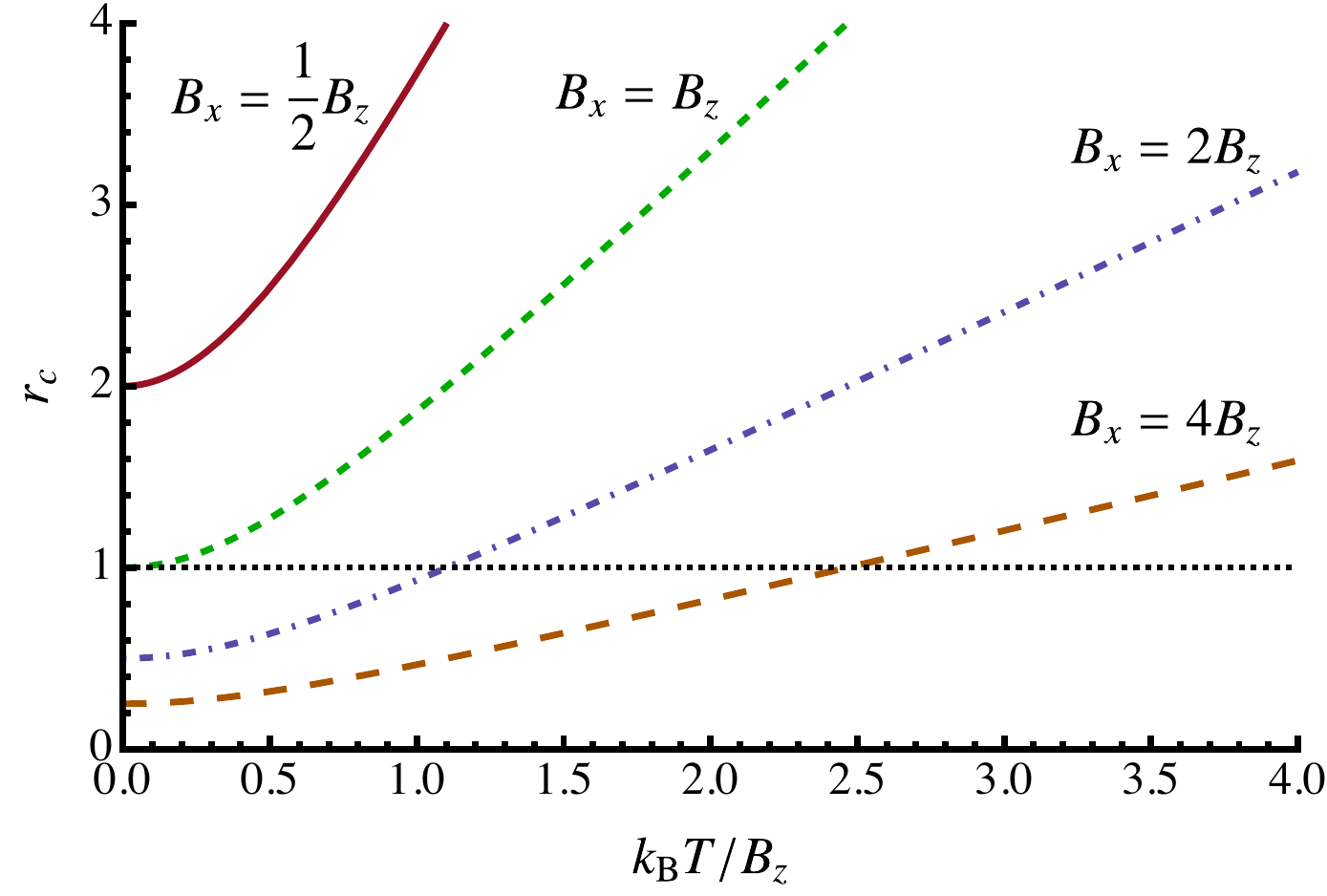}
\caption{Radius of convergence $r_c$ for $U$ and $F$ in the two-level system.
Different $B_x$ values show that divergent systems at $T=0$ will converge 
at $\lambda = 1$ for a sufficiently high $T$.}
\label{fig:RadConv}
\end{figure}

The closest pole to the origin corresponds to $n=0$, giving the radius of convergence 
\begin{equation}
\rc = \abs{\frac{B_z}{B_x} \sqrt{1 + \frac{\pi^2 \kB^2 T^2 }{4{B_z}^2}} }.
\end{equation} 
In the $T\rightarrow 0$ limit, this radius of convergence tends towards the expected value for the 
ground-state energy perturbation series and scales quadratically with respect to $T$ [see Figure~\ref{fig:RadConv}].
On the other hand, for $T\rightarrow\infty$, we recover the asymptotic behaviour $\rc \sim \frac{\pi \kB T}{2B_x}$. 
Therefore, the perturbation series will always converge at a sufficiently large $T$ as long as $B_x$ is finite.

\begin{figure*}[tb!]
\includegraphics[height=0.31\linewidth]{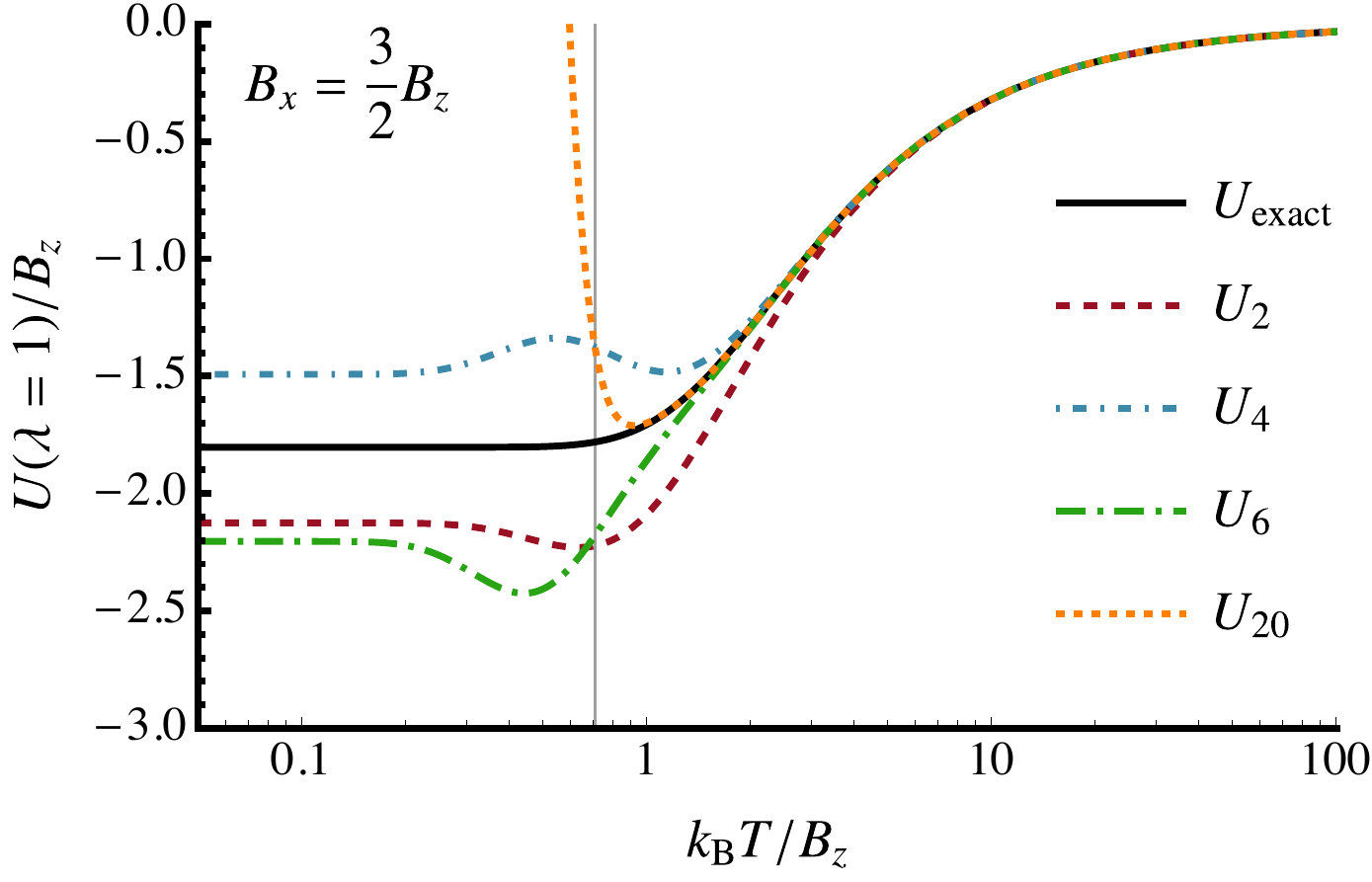}
\includegraphics[height=0.31\linewidth]{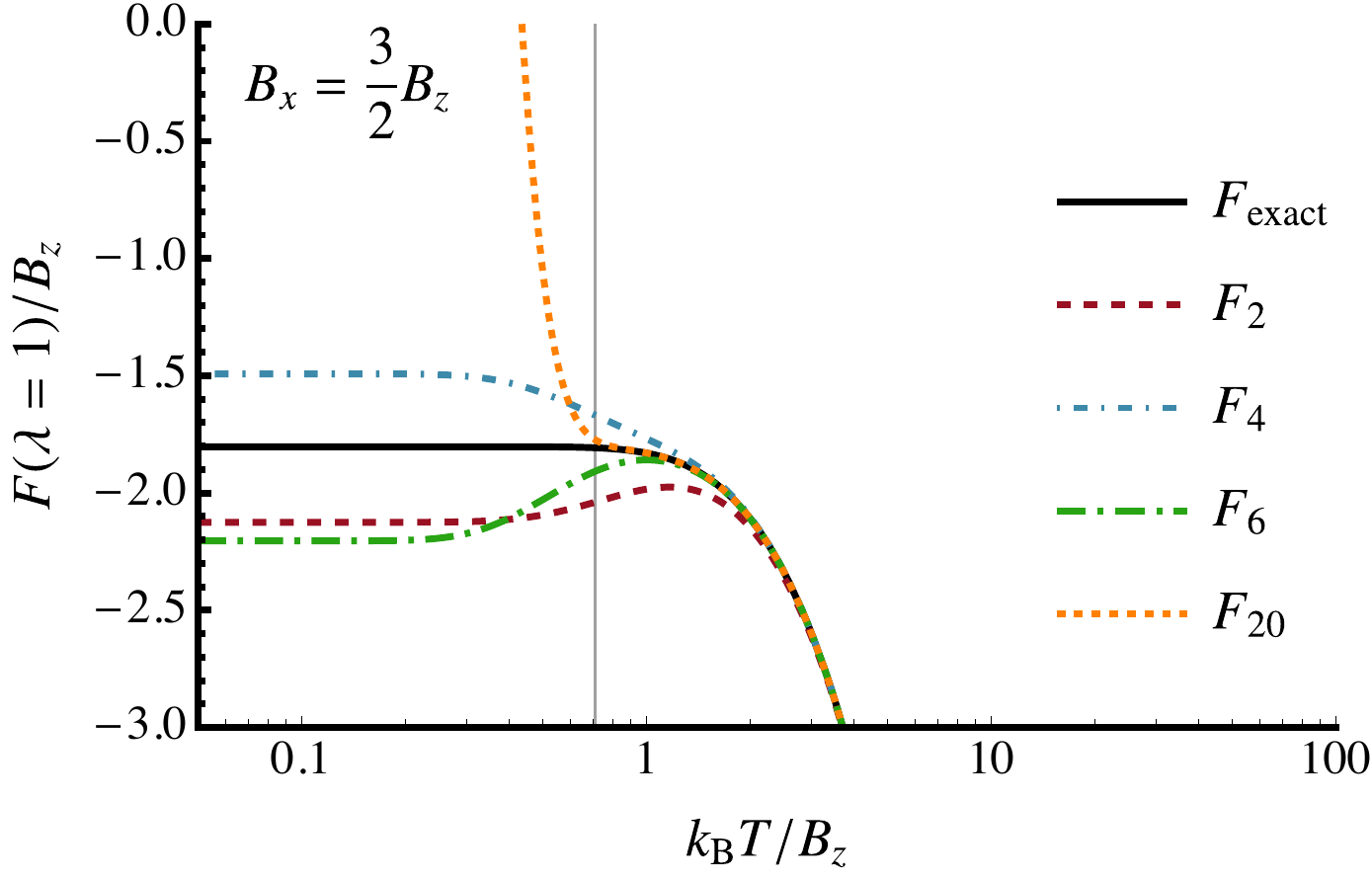}
\caption{Low-order perturbation expansions of the internal energy and the Helmholtz free energy for the 
two-level system with $B_x = \frac{3}{2}B_z$, evaluated at $\lambda =1$.
Here, $U_k$ or $F_k$ denote a partial sum of the perturbation expansion up to the $k$th-order. 
The vertical grey line at $\kB T=0.711\,763\, B_z$ indicates the lowest temperature where the perturbation 
expansions become convergent. }
\label{fig:TLtplot}
\end{figure*}

\red{Remarkably, we find that the finite-temperature perturbation series in the two-level system can converge at non-zero temperatures even when 
the zero-temperature expansion diverges.}
This scenario is illustrated in Figure~\ref{fig:TLtplot} for the internal energy and Helmholtz free energy 
with $B_x = \frac{3}{2} B_z$.
%
The temperature at which the thermodynamic perturbation expansion becomes convergent at $\lambda=1$ can be identified by solving 
$\rc > 1$, giving  $\kB T > \frac{2}{\pi}\sqrt{{B_x}^2 - {B_z}^2}$ in the two-level system.
For $B_x = \frac{3}{2} B_z$, the perturbation expansion becomes convergent at $\kB T=0.711\,763\, B_z$, as indicated by the 
vertical gray lines in Figure~\ref{fig:TLtplot}.
Numerical results indicating the accuracy of the low-order  corrections either side of this temperature are given in
Table~\ref{tab:numericalResults}. 
(Note that only terms of even order contribute to the expansion).
The numerical data in Figure~\ref{fig:TLtplot} and Table~\ref{tab:numericalResults} demonstrate that the Helmholtz free
energy and internal energy start to diverge at the same temperature, 
confirming that the zeros of the partition function (common to both thermodynamic potentials) are the origin of these divergences.

\begin{table}bb!]
\caption{Perturbation expansions of $U$ and $F$ for 
the two-level system with $B_x = \frac{3}{2}B_z$, evaluated at $\lambda = 1$.
Only terms with even order have a non-zero contribution. 
These series diverge at $\kB T = 0.5\, B_z$ and converge at $\kB T = 1.5\, B_z$.
}
\label{tab:numericalResults}
\begin{ruledtabular}
	\begin{ruledtabular}
		\begin{tabular}{ccccc}
           	&	\mc{2}{c}{$U(\lambda=1) / B_z$}	&	\mc{2}{c}{$F(\lambda=1) / B_z$} \\
																		\cline{2-3} \cline{4-5}
				Order	&	$\kB T = 0.5$	&	$\kB T = 1.5$	&	$\kB T = 0.5$	&	$\kB T= 1.5$	\\
			\hline
0		& $-0.96403$	&	$-0.58278$		&	$-1.00907$	&	$-1.35094$	\\
2		& $-2.20752$	&	$-1.73369$		&	$-2.09361$	&	$-2.00657$	\\
4		& $-1.34209$	&	$-1.42701$		&	$-1.57297$	&	$-1.91637$	\\
6		& $-2.40745$	&	$-1.53024$		&	$-2.02938$	&	$-1.93667$	\\
8		& $-0.96456$	&	$-1.49487$		&	$-1.55883$	&	$-1.93145$	\\
10	    & $-2.95966$	&	$-1.50701$		&	$-2.08118$	&	$-1.93289$	\\
12		& $-0.18759$	&	$-1.50284$		&	$-1.47585$	&	$-1.93248$	\\
\hline
Exact	&	$-1.80012$	&	$-1.50389$	&	$-1.80314$	&	$-1.93257$	\\
		\end{tabular}
	\end{ruledtabular}
\end{ruledtabular}
\end{table}

\red{%
On the other hand, it is also possible for a finite-temperature expansion to become divergent in the 
$T\rightarrow0$ limit while the corresponding zero-temperature expansion is convergent. 
In this scenario, there is a mismatch between the $T\rightarrow 0$ limit of $U^{(n)}$ and 
the corresponding zero-temperature ground-state correction $E_1^{(n)}$, known as the Kohn and Luttinger problem.\cite{Kohn1960} 
This issue originally suggested that the formulation of finite-temperature electronic perturbation theory may be incorrect, 
inspiring a comprehensive reformulation that expands all thermodynamic potentials on an equal footing.\cite{Hirata2013,Hirata2019,Hirata2020,Hirata2021b}
However, Hirata subsequently demonstrated that the Kohn--Luttinger problem still persists by
showing that $U^{(1)}$ and $U^{(2)}$ tend towards the wrong limit when the degeneracy is 
lifted at first order,\cite{Hirata2021}
\begin{subequations}
\begin{align}
\lim_{T\rightarrow0} U^{(1)} &= \mathbb{E}[E_k^{(1)}] \neq E_1^{(1)},
\\
\lim_{T\rightarrow0} U^{(2)} &=-\infty \neq E_1^{(2)},
\end{align}
\end{subequations}
where $\mathbb{E}[E_k^{(1)}]$ is the average of the first-order corrections for the degenerate reference states $k$.
These limits give a zero radius of convergence at zero temperature and Hirata concluded that the Kohn--Luttinger problem 
\textit{``originates from the non-analyticity of the Boltzmann factor at $T=0$''}.\cite{Hirata2021}
}

\red{%
Applying our analysis in the case of a degenerate reference state, we find that the zeros of the partition function must converge
onto $\lambda=0$ for $T\rightarrow0$.
Therefore, the zero-temperature continuum of poles in $U(\lambda)$ intersects $\lambda =0$ and the radius of convergence becomes zero, in agreement with Ref.~\onlinecite{Hirata2021}.
In contrast to zero-temperature perturbation theory, 
this feature arises from the properties of the partition function and only requires a 
degeneracy in the zeroth-order ground state;
it does not assume anything about the convergence properties of the zero-temperature eigenstates.
The presence of a non-analytic pole in $U(\lambda)$ at $\lambda=0$ then leads to an inherently divergent perturbation expansion.
Consequently, we can refine Hirata's concluding statement to say that the Kohn--Luttinger problem 
\textit{originates from a singularity in a thermodynamic potential
at $\lambda =0$ created by a zero of the partition function for $T\rightarrow0$}.
}

\red{However, this analysis does little to explain the origin of differences between the results of zero-temperature and finite-temperature perturbation
theory in the Kohn--Luttinger problem.
To elucidate this further, we fix $B_z=0$ in the two-level system to give degenerate reference states where the zero-temperature energies converge exactly at first-order, 
\begin{equation}
E_\pm(\lambda) = \pm  B_x \lambda.
\label{eq:deg_two-level}
\end{equation}
The corresponding internal energy is 
\begin{equation}
U(\lambda) = - B_x \lambda \tanh(\frac{B_x \lambda}{\kB T})
\end{equation}
with poles at
\begin{equation}
\lambda_\text{pole} = \pm \mathrm{i} \frac{\pi \kB T}{2B_x} (4n + 1) \quad \forall \quad n \in \mathbb{Z}.
\end{equation} 
The radius of convergence is then directly proportional to $T$ and
finite-temperature perturbation theory will diverge for $T\rightarrow0$, even though the zero-temperature energies are convergent.}

\begin{figure}[t!]
\includegraphics[width=\linewidth]{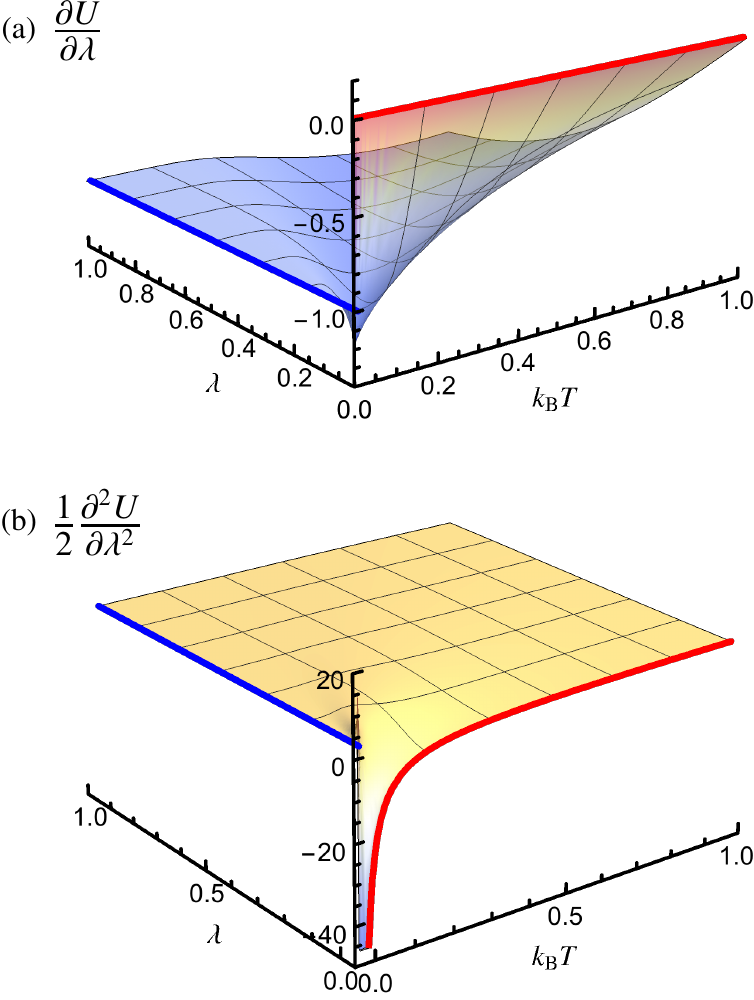}
\caption{(a) First-order and (b)  second-order contributions to the internal energy in the degenerate two-level system [Eq.~\eqref{eq:deg_two-level}] with $B_z=0$ and $B_x = 1$. 
Different values at $(\lambda,T)=(0,0)$ result from taking the $T\rightarrow0$ and $\lambda\rightarrow0$ limits in different orders, 
as illustrated by the red ($\lambda\rightarrow0$) and blue  ($T\rightarrow0$) curves.}
\label{fig:KLconundrum}
\end{figure}

\red{For this example, the derivatives required for the first- and second-order corrections to the internal energy are 
\begin{subequations}
\begin{align}
\frac{\partial U}{\partial \lambda} &= - \frac{{B_x}^2 \lambda}{\kB T} \sech^2\qty(\frac{B_x \lambda}{\kB T}) - B_x \tanh( \frac{B_x \lambda}{\kB T}),
\\
\frac{1}{2} \frac{\partial^2 U}{\partial \lambda^2} &= 
- \frac{{B_x}^2}{\kB T} \sech^2\qty(\frac{B_x \lambda}{\kB T})\, \qty[1-  \frac{{B_x}\lambda}{\kB T} \tanh(\frac{B_x \lambda}{\kB T})].
\end{align}
\end{subequations}
Both  functions are mathematically undefined at $\lambda = T =0$ and  the value obtained by perturbation theory will depend 
on the order in which this limit is approached.
For the first-order correction, two possible limits are
\begin{equation}
\lim_{T\rightarrow0^+} \qty( \lim_{\lambda \rightarrow 0^+} \frac{\partial U}{\partial \lambda} )  = 0
\quad\text{and}\quad 
\lim_{\lambda\rightarrow0^+} \qty( \lim_{T \rightarrow 0^+} \frac{\partial U}{\partial \lambda} ) = -B_x, 
\end{equation}
as illustrated in Figure~\ref{fig:KLconundrum}(a).
The first case corresponds to the zero-temperature limit of finite-temperature perturbation theory (red curve), recovering 
the average of the zero-temperature first-order corrections as described in Ref.~\onlinecite{Hirata2021}.
In contrast, the second case represents a perturbation expansion of the zero-temperature internal energy (blue curve) and 
recovers the expected behaviour $\lim_{T\rightarrow0} U^{(1)} = E_1^{(1)}$.
Similarly, the two limits for the second-order correction yield
\begin{equation}
\lim_{T\rightarrow0^+} \qty( \lim_{\lambda \rightarrow 0^+} \frac{1}{2}\frac{\partial^2 U}{\partial \lambda^2} )  = -\infty
\quad\text{and}\quad 
\lim_{\lambda\rightarrow0^+} \qty( \lim_{T \rightarrow 0^+} \frac{1}{2} \frac{\partial^2 U}{\partial \lambda^2} ) =0,
\end{equation}
as illustrated in Figure~\ref{fig:KLconundrum}(b).
Again, the first case recovers the divergent finite-temperature behaviour described in Ref.~\onlinecite{Hirata2021}
while the second case tends to the second-order correction of the exact ground state.
}

\red{These results suggest that the Kohn--Luttinger problem is fundamentally an order-of-limits issue that arises because the internal energy
with degenerate reference states is non-analytic for $\lambda=0$ and $T=0$ and the derivatives required for perturbation theory are mathematically undefined.
Because this situation arises from zeros of the partition function, the divergence depends on the degeneracy of the reference states regardless of
their coupling through the perturbation.
Therefore, this problem cannot be removed by simply using degenerate perturbation theory.\cite{Hirschfelder1974}
Instead, it appears that the only resolution is to remove the degeneracy using an alternative reference or 
partitioning, as recently proposed in Ref.~\onlinecite{Hirata2022}.
}

\red{Finally, having identified a relationship between branch cuts of the zero-temperature eigenstates and zeros of the partition function in the $T\rightarrow0$ limit}, a natural question is whether this concept can be extended to unify Lee--Yang theory with the convergence of exceptional points 
onto the real axis at quantum phase transitions.
Previous work by Cejnar and co-workers has investigated this connection by associating 
non-Hermitian degeneracies with point charges and connecting the position of these 
degeneracies in the complex plane with zeros of a Coulombic partition function.\cite{Cejnar2005,Cejnar2007}
However, we are not aware of any direct connection between Lee--Yang zeros and zero-temperature exceptional points at 
quantum phase transitions.

\begin{figure*}[htb!]
\includegraphics[width=\linewidth]{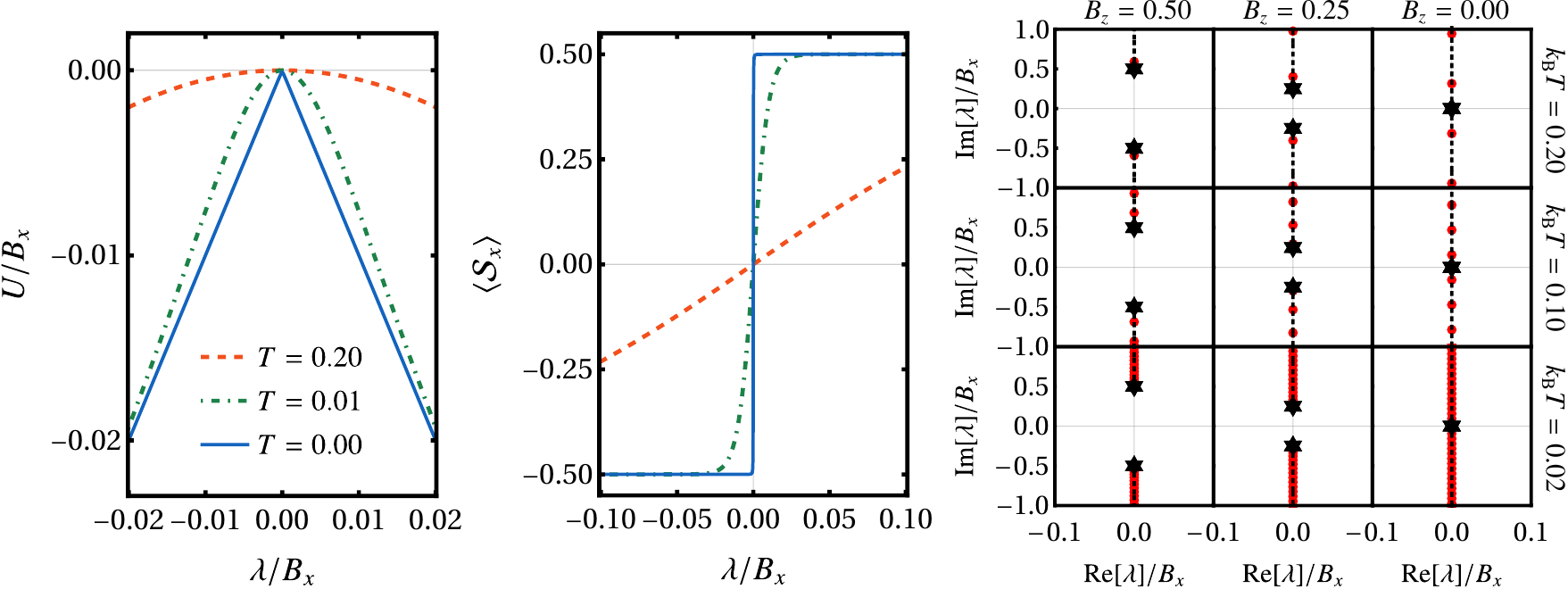}
\caption{A quantum phase transition occurs at $\lambda = 0$ in the two-level system with $T=0$ and $B_z =0$.
\textit{Left:} Internal energy at different temperatures, showing the quantum phase transition at $T=0$ as
a gradient discontinuity in the energy.
\textit{Middle:} Variation of the spin expectation value at the quantum phase transition.
\textit{Right:} Relationship between the exceptional points of the energy and zeros of the partition
function for various values of $B_z$ and $\kB T$ with $B_x = 1$.
}
\label{fig:EP_zeros}
\end{figure*}

The two-level spin-1/2 model provides a model of a quantum phase transition in the limit $B_z\rightarrow 0$. 
For a fixed value of $B_x > 0$, this quantum phase transition occurs at $\lambda = 0$, with the spin flipping its 
alignment from the negative-$x$ direction ($\lambda < 0$) to the positive-$x$ direction ($\lambda>0$).
This transition is demonstrated by a derivative discontinuity in the zero-temperature energy and a discontinuous jump in $\expval*{S_x}$
at $\lambda = 0$, as shown in Figure~\ref{fig:EP_zeros} (left and middle panels respectively). 
For $B_z \neq 0$, a complex-conjugate pair of exceptional points exist in the complex-$\lambda$ plane (Figure~\ref{fig:EP_zeros}: right panel),
creating an avoided level crossing on the real axis.
As expected, these zero-temperature exceptional points converge onto the real axis in the $B_z \rightarrow 0$ limit 
corresponding to the quantum phase transition.

This physical phase transition is mathematically equivalent to the analysis of finite-temperature perturbation theory in
the complex-$\lambda$ plane.
Therefore, we can use it to directly connect Lee--Yang zeros at quantum phase transitions with zero-temperature exceptional points.
Because the quantum phase transition only occurs when $T=0$ for $B_z =0$, we expect the zeros of partition 
function to converge onto the real axis in the $T \rightarrow 0$ and $B_z \rightarrow 0$ limit.
We have already shown that zeros of the 
partition function for $T\neq 0$ and $B_z \neq 0$ occur along the imaginary axis
extending out from the zero-temperature exceptional point.
For $T \rightarrow 0$, these zeros form a continuum along the branch cut that terminates at the exceptional point.
Therefore, in the $B_z \rightarrow 0$ limit, the convergence of the exceptional points onto the real axis 
also causes the zeros of the partition function to move closer to the real axis. 
The position of these zeros still has a complex component for $T\neq0$ and the internal energy is smooth along the real axis. 
However, when the zeros of the partition function converge onto the exceptional points for $T\rightarrow0$, they
must also converge onto the real axis as $B_z \rightarrow 0$, creating the expected behaviour for a quantum phase transition.

As a result, the convergence of Lee-Yang zeros and exceptional points onto the real axis arises from the 
same mathematics, providing a unified perspective on quantum phase transitions in the complex plane.
The exceptional point in the zero-temperature energy is simply the zero-temperature continuum limit of the poles in the 
internal energy caused by zeros of the partition function.
This result is a formal mathematical connection, in contrast to the previous Coulombic analogy between non-Hermitian 
degeneracies and zeros of the partition function.\cite{Cejnar2005,Cejnar2007}

In this work, we have shown that divergent finite-temperature perturbation theory is driven by poles created by Lee--Yang zeros 
of the partition function
that move further into the complex plane as the temperature increases.
Therefore, perturbation theory is more likely to be convergent at higher temperatures.
This result suggests that extrapolating analytic continuations of the perturbation corrections
at higher temperatures to $T=0$, in a similar approach to Refs.~\onlinecite{Mihalka2017,Surjan2018,Mihalka2019},
may provide improved accuracy when the zero-temperature expansion is divergent.
Alternatively, because the divergent perturbation expansions at $T\neq0$ arise from poles, the internal energy
will be particularly well-suited to resummation techniques based on Pad\'{e} approximants.\cite{BenderBook,Goodson2000,Goodson2012}
\red{For a degenerate reference, we find that the poles in the internal 
energy converge onto $\lambda=0$ in the $T\rightarrow0$ limit, creating a mathematically undefined 
perturbation expansion that is the origin of the Kohn--Luttinger problem.}

Finally, by extending our analysis to quantum phase transitions in the spin-1/2 model, we have 
demonstrated a direct mathematical connection between finite-temperature Lee--Yang zeros and zero-temperature exceptional points. 
This unifying connection provides a more complete picture of quantum phase transitions through the lens of complex analysis.
Recently, Lee--Yang zeros and exceptional points have been independently realised in 
experiments.\cite{Bittner2012,Doppler2016,Peng2015,Brandner2017} 
The possibility of experimentally probing the conversion of Lee--Yang zeros into an exceptional point at 
low temperatures is an exciting prospect for the discovery and verification of exotic non-Hermitian chemical physics.

\section*{Acknowledgements}
H.G.A.B.\ acknowledges Pierre-Fran\c{c}ois Loos for enlightening discussions in the early stages of this work and David Tew for supporting this research. H.G.A.B.\ was supported by New College, Oxford through the Astor Junior Research Fellowship. Y.S.\ was supported by a Summer Research Studentship from Hertford College, Oxford.

\section*{Author Declarations}
The authors have no conflicts to disclose.

\section*{Data Availability}

The data supporting the findings of this study will be made available in an open-access repository upon acceptance.

\appendix
\section{Analyticity of $Z(\lambda)$ near algebraic branch points}
\label{apdx:AnalyticU}
Consider the partition function $Z(\lambda)$ of 
an $N$-level system in the vicinity of an $k$-th order algebraic branch point $\lbp$ (where $k>1$). 
In this region, the states that are not involved in the branch point provide a smooth, analytic contribution $\mathcal{Z}(\lambda)$ and the partition function to be decomposed as
\begin{equation}
Z(\lambda)= \mathcal{Z}(\lambda) + \sum_{j=1}^{k} \exp(-\beta E_j(\lambda)),
\end{equation}
where the states indexed by $j$ coincide at the branch point.
To establish whether $Z(\lambda)$ is analytic at $\lbp$, we consider the first derivative 
\begin{equation}
\frac{\dd Z(\lambda)}{\dd \lambda}= \frac{\dd\mathcal{Z}(\lambda)}{\dd\lambda} 
- \beta \sum_{j=1}^{k} \frac{\dd E_j(\lambda)}{\dd \lambda} \exp(-\beta E_j(\lambda)).
\label{eq:Zdecomposed}
\end{equation}
Near the $k$th-order branch point, the corresponding energy levels behave as
\begin{equation}
E_j(\lambda) \approx E_{\text{bp}} + a(\lambda)^{\frac{1}{k}} \exp(\frac{2\pi \mathrm{i}j}{k})
\label{eq:Eexpansion}
\end{equation}
for some smooth real function $a(\lambda)$ with $a(\lbp) = 0$.
For example, in the two-level system considered in this letter, the energy levels near the branch point $E_{\pm}(\lambda) = \pm (B_z^2 + \lambda^2 B_x^2)^{1/2}$ correspond to 
$a(\lambda) = B_z^2 + \lambda^2 B_x^2$ and $k=2$.
The non-analyticity of the zero-temperature energy levels for $\lambda \rightarrow \lbp$ then arises
from the derivatives
\begin{equation}
\frac{\dd E_j(\lambda)}{\dd\lambda} =\frac{1}{k} \frac{ \dd a(\lambda)}{\dd \lambda}  a(\lambda)^{\qty(\frac{1}{k}-1)} \exp(\frac{2\pi \mathrm{i}j}{k}), 
\end{equation} 
where clearly the $a(\lambda)^{\qty(\frac{1}{k}-1)} $ term diverges in the limit $\lim_{\lambda \rightarrow \lbp} a(\lambda) = 0$ for $k>1$.
Inserting Eq.~\eqref{eq:Eexpansion} into Eq.~\eqref{eq:Zdecomposed} and considering the $\lambda \rightarrow \lbp$ limit then yields
\begin{align}
\frac{\dd Z(\lambda)}{\dd \lambda} \approx 
&\frac{\dd\mathcal{Z}(\lambda)}{\dd\lambda} 
\label{eq:approxZ}
\\
&- \beta \frac{1}{k} \frac{ \dd a(\lambda)}{\dd \lambda}  a(\lambda)^{\qty(\frac{1}{k}-1)} \exp(-\beta \,E_\text{bp}) 
\underbrace{\sum_{j=1}^{k} \exp(\frac{2\pi \mathrm{i} j}{nk})}_{0} .
\nonumber
\end{align}
Because $\sum_{j=1}^{k} \exp(\frac{2\pi \mathrm{i} j}{nk})=0$, we find that $\frac{\dd Z(\lambda)}{\dd \lambda}$ remains finite at the branch point even though the zero-temperature energy levels are non-analytic.
This derivation can be trivially extended to show that high-order derivatives of the partition function
are also finite at a branch point.
Consequently, the partition function is a complex analytic function of $\lambda$, even in the presence of non-analytic algebraic branch points in the zero-temperature energy levels.
A similar approach can be applied to the numerator of the internal energy.

\section*{References}
\bibliography{manuscript}

\end{document}